\documentclass[aps,prl,twocolumn, showpacs,superscriptaddress]{revtex4-1}

\usepackage{graphicx}
\usepackage{dcolumn}
\usepackage{bm}
\usepackage{epsf}
\usepackage{epsfig}
\usepackage{amssymb}
\usepackage{verbatim}
\usepackage{amsmath}
\usepackage{longtable}
\usepackage[utf8]{inputenc}
\usepackage{helvet}
\usepackage[english]{}
\usepackage{subdepth}

\begin{document}

\title{Observation of double pygmy resonances in $^{195,196}$Pt\\and enhanced astrophysical reaction rates}

\author{F.~Giacoppo}
\email[]{francesca.giacoppo@fys.uio.no}
\affiliation{Department of Physics, University of Oslo, N-0316 Oslo, Norway}

\author{F.L.~Bello Garrote}
\affiliation{Department of Physics, University of Oslo, N-0316 Oslo, Norway}

\author{T.K.~Eriksen}
\affiliation{Department of Physics, University of Oslo, N-0316 Oslo, Norway}

\author{A.~G\"{o}rgen}
\affiliation{Department of Physics, University of Oslo, N-0316 Oslo, Norway}

\author{M.~Guttormsen}
\affiliation{Department of Physics, University of Oslo, N-0316 Oslo, Norway}

\author{T.W.~Hagen}
\affiliation{Department of Physics, University of Oslo, N-0316 Oslo, Norway}

\author{A.C.~Larsen}
\affiliation{Department of Physics, University of Oslo, N-0316 Oslo, Norway}

\author{B.V.~Kheswa}
\affiliation{Department of Physics, University of Oslo, N-0316 Oslo, Norway}
\affiliation{Department of Physics, University of Stellenbosch, 7602 Stellenbosch, South Africa}

\author{M.~Klintefjord}
\affiliation{Department of Physics, University of Oslo, N-0316 Oslo, Norway}

\author{P.E.~Koehler}
\affiliation{Department of Physics, University of Oslo, N-0316 Oslo, Norway}

\author{H.T.~Nyhus}
\affiliation{Department of Physics, University of Oslo, N-0316 Oslo, Norway}

\author{T.~Renstr{\o}m}
\affiliation{Department of Physics, University of Oslo, N-0316 Oslo, Norway}

\author{E.~Sahin}
\affiliation{Department of Physics, University of Oslo, N-0316 Oslo, Norway}

\author{S.~Siem}
\affiliation{Department of Physics, University of Oslo, N-0316 Oslo, Norway}

\author{T.G.~Tornyi}
\affiliation{Department of Physics, University of Oslo, N-0316 Oslo, Norway}
\affiliation{Institute of Nuclear Research of the Hungarian Academy of Sciences, H-4001 Debrecen, Hungary}

\date{\today}
\begin{abstract}
Our measurements of $^{195,196}$Pt $\gamma$-strength functions show a double-humped enhancement in the $E_{\gamma}= 4-8$ MeV region. For the first time, the detailed shape of these resonances is revealed for excitation energies in the quasicontinuum.
We demonstrate that the corresponding neutron-capture cross sections and astrophysical reaction rates are increased by up to a factor of 2 when these newly observed pygmy resonances are included. These results lend credence to theoretical predictions of enhanced reaction rates due to such pygmy resonances and hence are important for a better understanding of r-process nucleosynthesis.
\end{abstract}


\maketitle

The phenomenon of resonances appears in a great variety of systems in nature. An example from macroscopic systems is the Tacoma Bridge in  USA  which collapsed as a consequence of unstable oscillations driven by strong wind blasts~\cite{Tacoma_Bridge}. A microscopic resonance effect is, for instance, the coherent laser light generated in a resonator cavity~\cite{laser}. In classical systems, the resonance frequency is the pivotal parameter, while in a quantum-mechanical system the energy is the equivalent of the classical frequency, as the energy is given by $E = \hbar \omega$. 

The most dominant and best known resonance observed in atomic nuclei is the isovector giant electric dipole resonance (GEDR)~\cite{Harakeh}, originating from the oscillation of the proton and neutron clouds, against each other, in connection with the absorption or emission of $\gamma$ rays~\cite{Migdal}. The GEDR has been experimentally investigated in most of the stable nuclei and has recently been established as a general feature also for several neutron-rich isotopes~\cite{132Sn_GSI}.

Fifty years ago Bartholomew and his co-workers observed a systematic enhancement in the region between 5 and 7 MeV of the electromagnetic response of many nuclei exposed to thermal neutrons (see for instance Ref.~\cite{Bartholomew_ncapture}). This excitation mode, which was later denoted as the pygmy dipole resonance (PDR), drew a lot of attention and stimulated efforts aiming at disclosing its nature. Currently, the PDR is accepted to be closely related to the vibration of the loosely bound protons or neutrons against the neutron-proton core. However, this collective interpretation is not sufficient to explain the experimentally observed low-energy dipole strengths, and the coupling to more complex configurations should also be taken into account~\cite{PDR_review}. In the past decades other excitation modes have been predicted and observed as well, such as the M1 scissors resonance~\cite{Heyde_M1, Magne_PRL}. 

The reduced average $\gamma$-decay probability, which is called the $\gamma$-strength function ($\gamma$SF), is directly connected to the nucleus' ability to emit or absorb a photon of a specific energy. Resonances in the $\gamma$SF give a higher probability for $\gamma$ emission or absorption. 
 
Platinum isotopes close to stability are peculiar in their nuclear level structure and electromagnetic properties, as they belong to the Os-Pt-Hg transition region between strongly deformed and spherical nuclei. In experimental and theoretical studies they are described as triaxial, $\gamma$-soft nuclei with an average $\gamma\approx30^o$ and a dynamic shape~\cite{gammasoft_195Pt, gammasoft_196Pt}.
The element Pt is also of special interest from an astrophysical point of view, as it is predominantly produced in rapid neutron-capture \mbox{($r$-)} processes in extremely neutron-rich astrophysical environments that are still to be uniquely identified~\cite{Arnould}. Moreover, there is an r-process peak in the solar-system element abundance at $A \approx 195$. Depending on the physical conditions at the r-process site, nuclear properties such as resonances in the $\gamma$SF could significantly change the neutron-capture reaction rates, with possible implications for the final calculated abundances~\cite{Goriely}.

In this Letter, we present new measurements of the $\gamma$SF of $^{195,196}$Pt. For the first time, the $\gamma$SFs of such $\gamma$-soft nuclei have been measured for a wide range of excitation energies in the quasicontinuum and for the full $\gamma$-energy range of 1 MeV $\lesssim E_{\gamma}\leqslant S_n$, that is, indeed, the most relevant energy region for astrophysical applications. In the following, particular emphasis will be placed on the strong  and very similar double pygmy structure observed in both Pt nuclei in the vicinity of the neutron separation energy. Moreover, calculated ($n,\gamma$) cross sections of importance for neutron-capture reaction rates are also presented. 

The experiments were carried out at the Oslo Cyclotron Laboratory, where it was possible to determine, simultaneously, the nuclear level density and the $\gamma$SF of Pt isotopes, applying the Oslo method~\cite{Schiller, Larsen}. A deuteron and a proton beam accelerated to 11.3 MeV and 16.5 MeV, respectively, were used to irradiate a self-supporting $^{195}$Pt target enriched to 97.3\% and with mass thickness 1.5(2) mg/cm$^2$. The Silicon Ring (SiRi) particle detector system~\cite{SiRi} and the CACTUS multidetector array~\cite{CACTUS} measured particle-$\gamma$ coincidences with a time resolution of $\Delta t\approx$15 ns. SiRi consists of 64 silicon ${\Delta}E$ and $E$ telescopes with a thickness of 130 and 1550 ${\mu}$m, respectively, covering 8 angles from $\theta$=$126^{o}$ to $\theta$=$140^{o}$ in the backward direction with respect to the beam axis. A set of 28 collimated $5$"$\times5$" NaI(Tl) $\gamma$-ray detectors forms the CACTUS array. Its solid angle coverage is 16\% and its total detection efficiency is 15.2(1)$\%$ at E$_{\gamma}$= 1332~keV.

From each particle-$\gamma$ coincidence event it was possible to extract the excitation energy \textit{E} of the residual nucleus based on particle identification and the reaction kinematics. The measured $\gamma$ rays were hence tagged for each excitation energy bin. The first stage of the Oslo method is an unfolding procedure, where the spectra of emitted $\gamma$ rays were corrected by means of the CACTUS response functions~\cite{Unfolding}. 
\begin{figure}[t]
\includegraphics[width=\columnwidth]{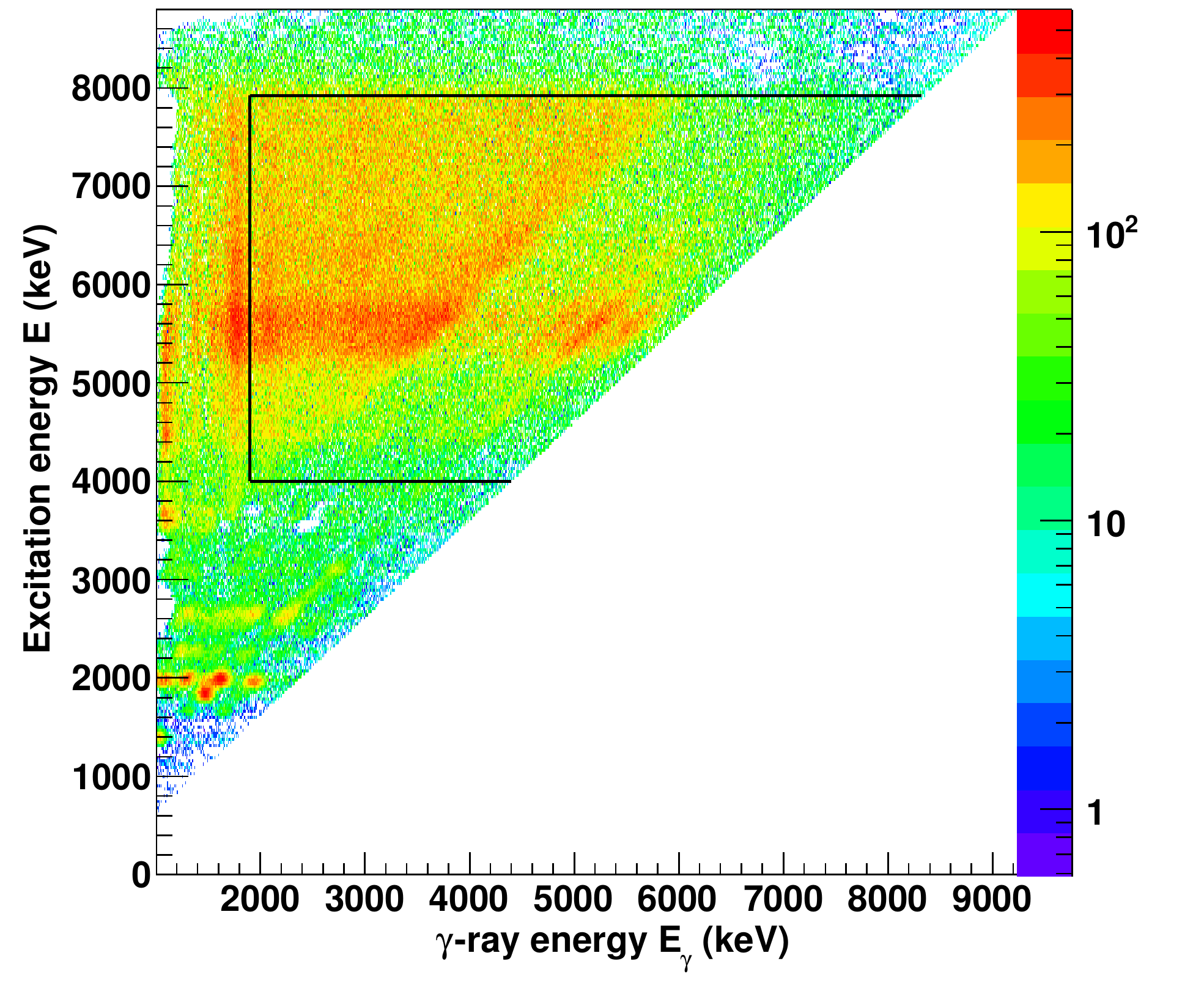}    
\caption{(Color online) Excitation energy vs primary $\gamma$-ray energy of $^{196}$Pt. The black lines define the region of data chosen for the extraction of the level density and $\gamma$SF. \label{FG_matrix}}
\end{figure}
Further, the distribution of the primary (first generation) $\gamma$ transitions for each excitation-energy bin was drawn out through an iterative subtraction technique~\cite{Primarygen}. 
Finally, the matrix of the first-generation $\gamma$ rays tagged on excitation energy ${P}(E, E_{\gamma})$ was generated as shown in Fig.~\ref{FG_matrix} for $^{196}$Pt. 

According to the Fermi's golden rule~\cite{Fermi}, \textit{P} can be factorized in the product of the level density at the final excitation energy $\rho(E-E_{\gamma})$ and the transition matrix element between the initial and the final states. The latter is related to the $\gamma$-transmission coefficient $\mathcal{T}(E\rightarrow E-E_{\gamma})$ which is assumed to be independent of the excitation energy, as given by the Brink-Axel hypothesis~\cite{Brink, Axel}:
\begin{equation}
P(E, E_{\gamma})\propto\rho(E-E_{\gamma})\mathcal{T}(E_{\gamma}).
 \label{eq-1}
\end{equation}
Eq.~(\ref{eq-1}) is valid for a statistical decay  process, i.e.~when the nucleus has thermalized and formed a compound state before $\gamma$ decay~\cite{BohrMottelson}. 
Therefore, we only consider the statistical region in the primary $\gamma$-ray matrix (see Fig.~\ref{FG_matrix}).

Through a least chi-square minimization procedure, the functional form of the level density and the $\gamma$-transmission coefficient can be deduced from ${P}(E, E_{\gamma})$~\cite{Schiller}. In order to obtain the absolute distributions of $\rho$ and $\mathcal{T}$, these two functions are normalized to independent experimental data. In particular $\rho$ is corrected to the density of known levels at low excitation energy~\cite{NNDC} and to the level density at the neutron separation energy $\rho(S_n)$ that is derived from the average neutron resonance spacing $D_0$~\cite{Paul_Pt_astro}. We assume that the spin distribution follows the energy-dependent formula proposed in Ref.~\cite{E&B2009}. The correction parameter for $\mathcal{T}$ is related to the average total radiative width $\left<\Gamma_{\gamma}\right>$ at $S_n$~\cite{Paul_Pt_astro}.
Further details on the normalization procedure are found in Refs.~\cite{Schiller, Larsen}.

\begin{figure}[t]
\includegraphics[width=\columnwidth]{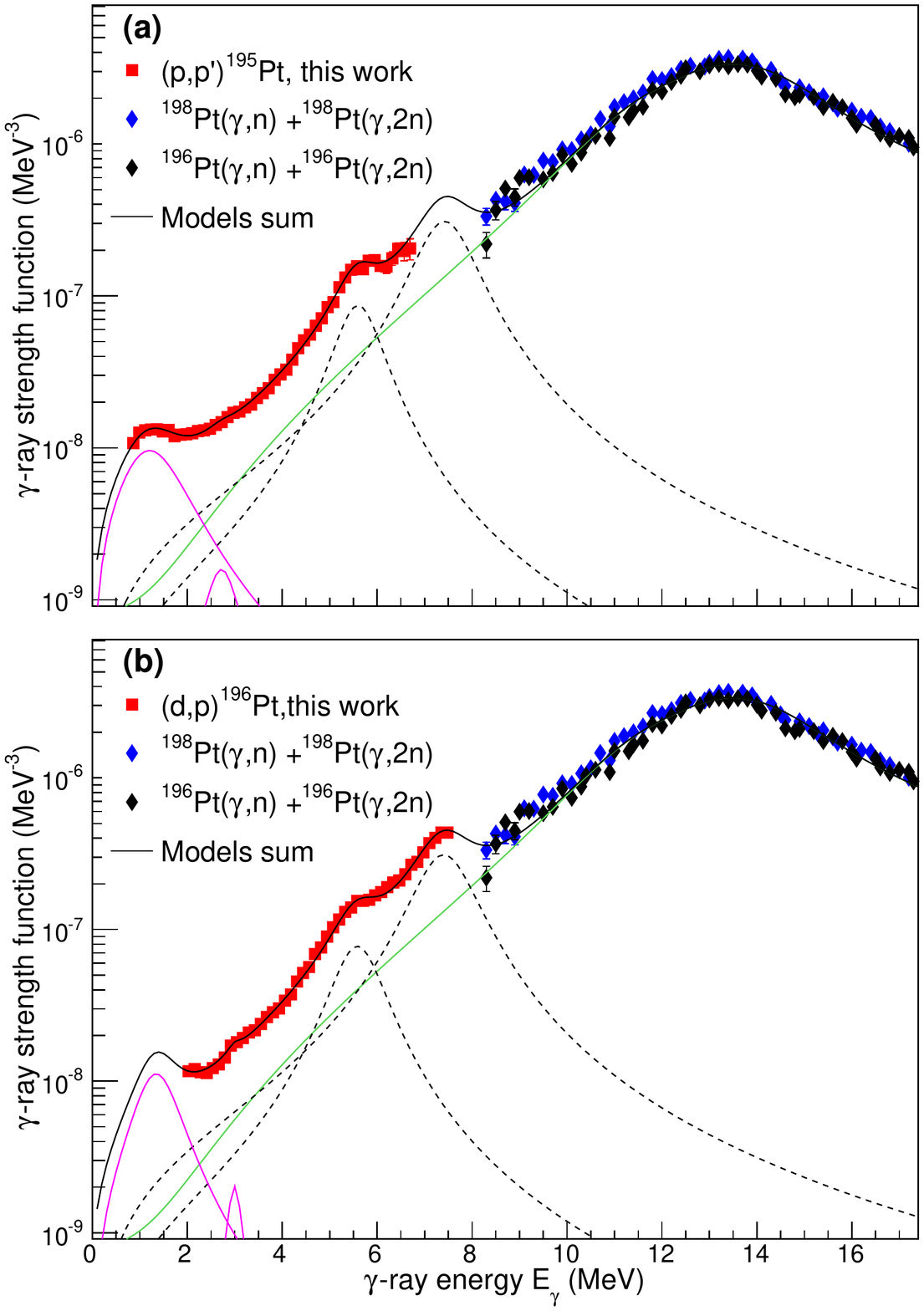}
\caption{(Color online) Gamma strength function of (a) $^{195}$Pt and (b) $^{196}$Pt from the present work (red squares) together with data from photonuclear reactions (black and blue diamonds~\cite{Goryachev}).The EGLO and SLO models are here applied to best fit the GEDR (green line) and PDR (dotted black lines) respectively. Two additional SLO (violet lines) are included to reproduce the strength at low energy. The solid black lines give the sum of the models.\label{strengths_195_196Pt}}
\end{figure}

In the quasicontinuum, the dipole transitions play a dominant role~\cite{RIPL3}. Therefore, the $\gamma$SF can be calculated from the  transmission coefficient via the relation: 
\begin{equation}
f(E_{\gamma})=\frac{1}{2\pi} \frac{\tilde{\mathcal{T}}(E_{\gamma})}{E_{\gamma}^3}.
\end{equation}
The resulting $\gamma$SFs are presented in Fig.~\ref{strengths_195_196Pt}, together with photonuclear data  which cover the energy region above $S_n$ where the GEDR has its maximum~\cite{Goryachev}. Here the dipole strength is well reproduced by the enhanced generalized Lorentzian (EGLO) model~\cite{EGLO, RIPL3}. In order to take into account the vibration of the nucleus along each of the principal axes $k=$1,2,3, a combination of three components is chosen, with free parameters $E_k$, $\Gamma_k$ and cross section $\sigma_k$ as listed in Table~\ref{GEDR_model}. The splitting of the centroid energy $E_k$ and the spreading in the width $\Gamma_k$ are related to the details of the nuclear shape, i.e. the deformation $\mathit{\beta}$ and triaxiality $\mathit{\gamma}$~\cite{Alhassid}. The EGLO model is also rather successful in reproducing experimental data in the low $\gamma$-energy region with a non-vanishing limit depending on the nuclear temperature $T_f$ of the final states~\cite{KMF}. Consistently with the Brink-Axel hypothesis, $T_f$ is assumed to be a constant parameter which is fitted to the experimental data.

The $\gamma$SFs presented in this work cover the energy region below the neutron separation energy. However, for $^{195}$Pt it was possible to extend the analysis to about 300 keV above $S_n$ by selecting the proper data subset in the first-generation matrix.

For both Pt isotopes a double-humped structure is observed on the tail of the GEDR. A recent ($\gamma$,$\gamma'$) experiment aiming at measuring the ground-state photo-absorption cross section of $^{196}$Pt, displays two resonances in the same energy region as in the present work~\cite{Pt_Dresden}. With our new data, we show that these bumps are present also for the $\gamma$-decay channel of excited states in the quasicontinuum. As shown in Fig.~\ref{strengths_195_196Pt} the resonances are fitted with standard Lorentzian (SLO) functions~\cite{RIPL3}, the corresponding parameters are listed in Table~\ref{Pygmy_model}. 
\begin{table}[t]  
 \caption{Parameter values used for the theoretical description of the GEDR in $^{195,196}$Pt.\label{GEDR_model}}
\begin{ruledtabular}
 \begin{tabular}{l*{4}{c}}
             &$E_k$		&$\Gamma_k$		&$\sigma_k$		&$T_f$			\\
  	    & (MeV)		&(MeV)			&(mb)			&(MeV) 			\\
\hline
First      &12.42		&4.57			& 160.0			&0.12			\\
Second  &13.62		& 3.44			& 245.0 			&0.12			\\
Third     &14.54		& 3.70			& 200.0 			&0.12			\\
\end{tabular}
\end{ruledtabular}
 \end{table}
\begin{table}[t]  
 \caption{Parameter used for the models of PDR in $^{195,196}$Pt.\label{Pygmy_model}}
\begin{ruledtabular}
 \begin{tabular}{l*{6}{c}}
		 &$E_{1}$ 		&$\Gamma_{1}$		&$\sigma_{1}$		&$E_{2}$ 			&$\Gamma_{2}$ 		&$\sigma_{2}$	       \\
   		 &(MeV)		&(MeV)			&(mb)			&(MeV) 			&(MeV)			&(mb)   		       \\
\hline
$^{195}$Pt    &5.62		&1.05 			& 5.6 			&7.45			&1.35			&26.5 		        \\
$^{196}$Pt	   &5.61		&1.15			& 5.0 			&7.44			& 1.40			&26.6 		        \\
\end{tabular}
\end{ruledtabular}
\end{table}

The average energy of the two peaks is consistent with a $A^{-1/3}$ law, characteristic of  M1 spin-flip excitations~\cite{Richter}. Within this context the splitting of the strength into two major components could be explained in terms of the nuclear deformation. However, the integrated strength $\sum B(M1)$ would be approximately a factor of ten larger than in closed-shell nuclei~\cite{Heyde_M1, M1_Pb}. In addition, there is evidence of E1 transitions in the proximity of the neutron separation threshold from average neutron resonance data~\cite{Mughabghab}. If one assumes that all the excess of strength is E1, this yields to 0.61\% and 1.3\% of the Thomas-Reiche-Kuhn (TRK) sum rule for $3.5\leqslant E_{\gamma}\leqslant6.7$ MeV and $3.5\leqslant E_{\gamma}\leqslant7.5$ MeV in $^{195,196}$Pt, respectively.
According to the model parametrization chosen in this work, the total contribution of the two pygmies exhausts about 2.3\% of the TRK sum rule in both nuclei. It is interesting to observe that this collective mode has the same intensity in the two adjacent isotopes under study. 

Unfortunately, the present experimental technique is not sensitive to the electromagnetic character of the $\gamma$ transitions; therefore the nature of the observed double-humped enhancement is not yet firmly established.
 \begin{figure}[t]
 \begin{center}
\includegraphics[clip,width=\columnwidth]{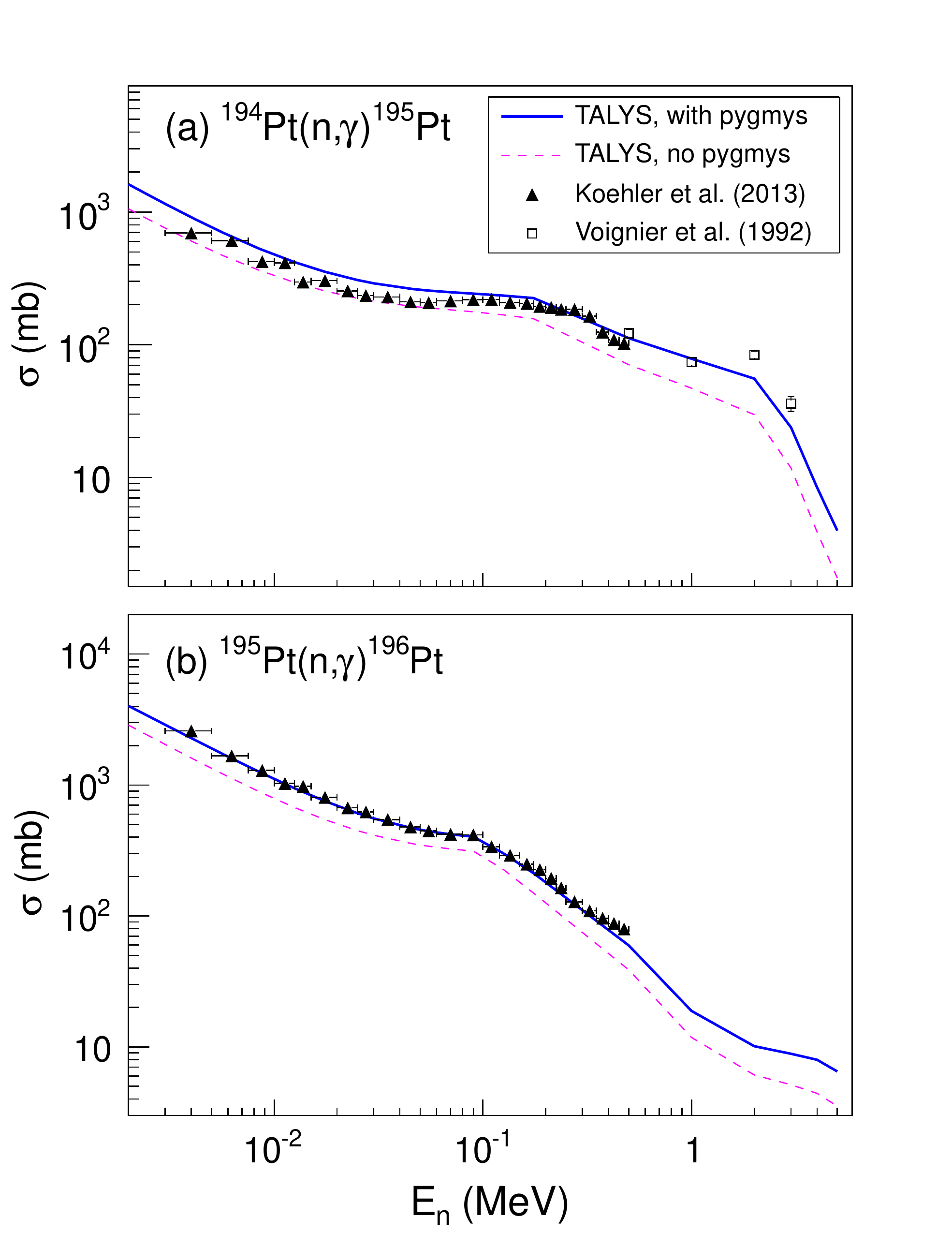}
 \caption{(Color online) Neutron-capture cross sections for (a) $^{194}$Pt($n,\gamma$)$^{195}$Pt and (b) $^{195}$Pt($n,\gamma$)$^{196}$Pt.
 The blue, solid line is the calculated cross section including the two pygmies; the magenta, dashed line is the result without the pygmies.
 Data are from Ref.~\cite{Paul_Pt_astro} (black triangles) and~\cite{voignier1992} (open squares).}
 \label{fig:xsec}
 \end{center}
 \end{figure}
One should also mention that a weaker broad excess of strength is observed in the low-energy region $E_{\gamma}<$3.5 MeV of the $^{195}$Pt $\gamma$SF and a similar behavior cannot be excluded for $^{196}$Pt (see Fig.~\ref{strengths_195_196Pt}). According to theoretical predictions, this is the energy domain where the M1 scissors resonance is expected. A nuclear resonance fluorescence experiment has revealed similar strength for $^{196}$Pt~\cite{vonBrentano}.

To investigate the impact of the pygmy resonances on astrophysical ($n,\gamma)$ reaction rates, 
we have performed cross-section and rate calculations using the reaction code TALYS~\cite{TALYS} 
for the radiative neutron-capture reactions $^{194,195}$Pt($n,\gamma$)$^{195,196}$Pt.
One of the ingredients is the level density, which has been described by the constant-temperature model~\cite{CT_model} using the parameterization from~\cite{E&B2009}. In particular the temperature $T = 0.63$ MeV has been chosen for both nuclei,
and energy shifts of $E_0 = -2.07$ MeV and $-0.81$ MeV for $^{195,196}$Pt, respectively.
These level densities describe our data very well above the pair-breaking region ($E> 2\Delta \approx 2$ MeV). The parameters for
the $\gamma$-strength function are given previously in the text. We have here assumed that the pygmy at $E_\gamma \approx 5.6$ 
MeV is of $M1$ type while the one at $E_\gamma \approx 7.5$ MeV is of $E1$ type, although tests have verified that a swapping of 
the electromagnetic characters will not significantly change the cross sections (a relative difference of maximum 6\% is found). 
For the neutron optical-model potential
we applied the phenomenological parameterization of Koning and Delaroche~\cite{koning2003} as implemented in TALYS, 
but with adjustments of the parameters $a_V$ and $a_D$ according to the findings of Ref.~\cite{Paul_Pt_astro}.

The resulting calculated cross sections are displayed in Fig.~\ref{fig:xsec}. We find that the calculations 
describe the directly measured ($n,\gamma$) cross sections of Refs.~\cite{Paul_Pt_astro,voignier1992} very well, especially for 
($n,\gamma$)$^{196}$Pt where there is an excellent agreement with the data.
It is clear that the two pygmies have a significant impact on the ($n,\gamma$) cross sections, giving an increase up to $\approx 85$\%
and $\approx 230$\%
for 5 MeV incoming neutrons for ($n,\gamma$)$^{195,196}$Pt, respectively.

\begin{figure}[bt]
 \begin{center}
\includegraphics[clip,width=8.3cm, height=13.5cm]{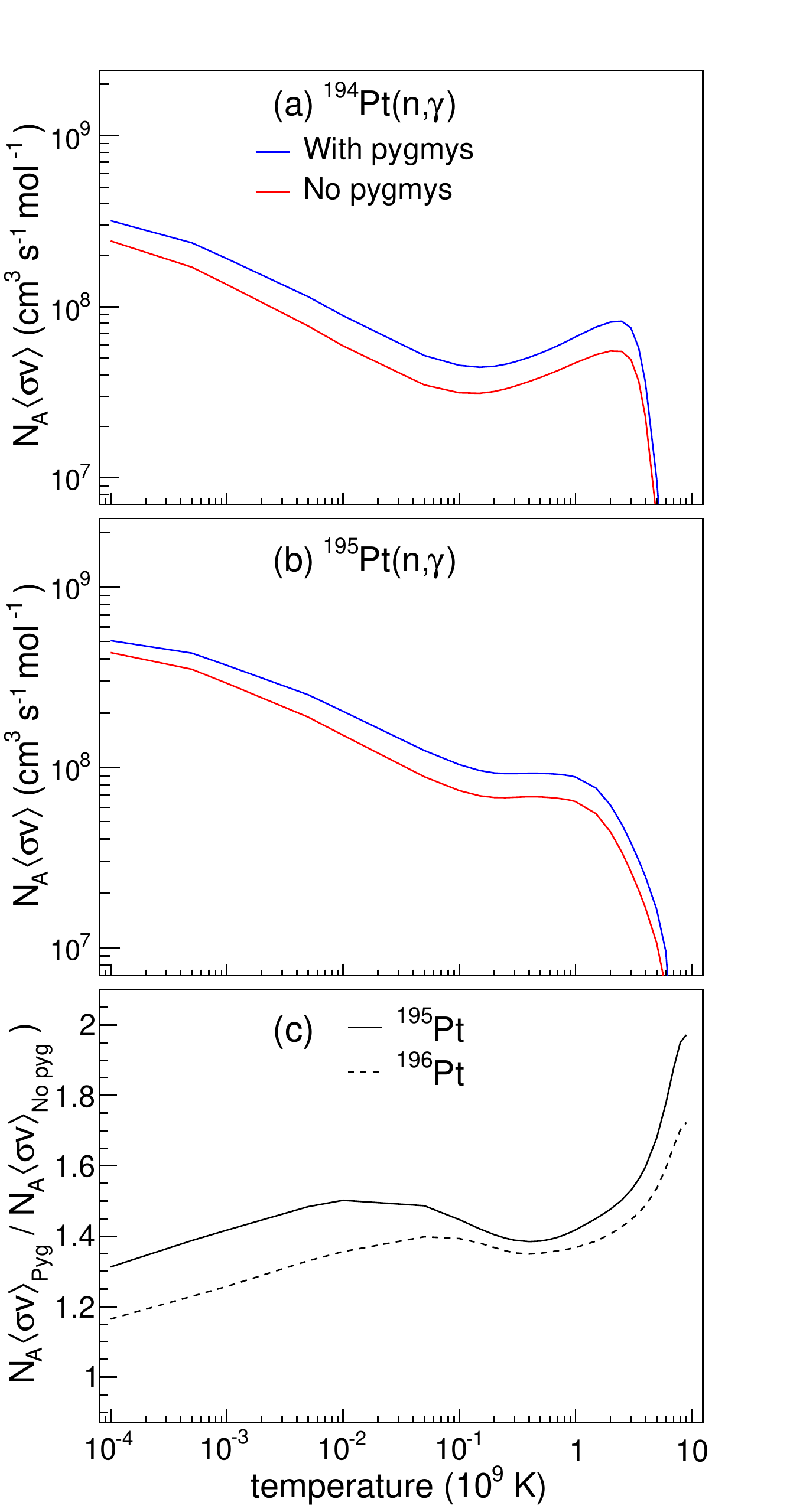}
 \caption{(Color online) Maxwellian-averaged reaction rates for (a) $^{194}$Pt($n,\gamma$)$^{195}$Pt and (b) $^{195}$Pt($n,\gamma$)$^{196}$Pt.
 The blue, solid line is the rate including the two pygmies; the red line is the result without the pygmies.
 In panel (c), the ratio of the reaction rates with and without the pygmies are shown for $^{195,196}$Pt (solid and dashed line, respectively).}
 \label{fig:rates}
 \end{center}
 \end{figure}

We have also calculated astrophysical ($n,\gamma$) reaction rates as a function of temperature for $^{195,196}$Pt again using TALYS. These results are displayed in Fig.~\ref{fig:rates}: clearly the inclusion of the pygmy resonances enhance the rates significantly.
As shown in Ref.~\cite{Goriely}, such  resonance structures in the vicinity of the neutron threshold may have profound implications for the heavy-element nucleosynthesis. As the Pt isotopes are located at the $A \approx 195$ $r$-process peak, the experimental confirmation of such structures in this region is essential to obtain a reasonable accuracy of the relevant ($n,\gamma$) reaction rates. It is therefore highly desirable to perform experiments with very neutron-rich nuclei in the Pt mass region to investigate whether these pygmies are present in the $\gamma$SF and if so, determine their intensities. 

To summarize, in this Letter we have shown the occurrence of two resonance-like structures close to the neutron separation energy 
in $^{195,196}$Pt. It is demonstrated that these resonances exhaust a significant part of the total $\gamma$-decay strength. 
Using our data on the level densities and $\gamma$SFs as input, together with information from recent neutron-resonance measurements
at ORELA, we find a very good agreement with radiative neutron-capture cross sections for these nuclei. The resulting
astrophysical ($n,\gamma$) reaction rates are found to increase up to a factor of $\approx$ 2 when the pygmy resonances are
included in the $\gamma$SFs. The impact of low-lying pygmy resonances in the estimation of neutron-capture cross sections and rates has been foreseen by theoretical calculations and it is confirmed experimentally in this Letter for the first time.

For the fundamental understanding of these structures, it is crucial to nail down the electromagnetic character of both resonances.
This will be subject for future experiments. Also, to confirm or disprove the existence of these resonances for very neutron-rich nuclei
in this mass region, radioactive-beam experiments are envisaged. The results from such experiments will be indispensable
for the understanding of the $\gamma$SF in exotic nuclei, and accordingly of their astrophysical applications.

We would like to thank J.C. M{\"u}ller, A. Semchenkov and J. Wikne at the Oslo Cyclotron Laboratory for providing stable proton and deuteron beams during the experiments. This work was supported by the Research Council of Norway.


\end{document}